\documentclass{article}

\usepackage{times}
\usepackage[square]{natbib}
\usepackage{url}
\usepackage{epstopdf}
\usepackage[pdftex]{graphicx}
\usepackage{newlfont}    
\usepackage{amsmath}

\newcommand{\be}{\begin{equation}}
\newcommand{\ee}{\end{equation}}
\newcommand{\ba}{\begin{eqnarray}}
\newcommand{\ea}{\end{eqnarray}}

\begin{document}


\begin{center}
{\Huge{Statistical Physics Approaches to Seismicity}} \\
\vspace{0.5cm}
{\Large{ \textsc{ Didier Sornette$^1$ and Maximilian J. Werner$^2$} } } \\
\vspace{0.3cm} 
$^1$ Department of Management, Technology and Economics, ETH Zurich\\
$^2$ Swiss Seismological Service, Institute of Geophysics, ETH Zurich \\
\end{center}

\tableofcontents
\section{Glossary}

\begin{itemize}

\item \textbf{Chaos}: Chaos occurs in dynamical systems with two ingredients: (i) nonlinear recurrent re-injection of the dynamics into a finite domain in phase space and (ii) exponential sensitivity of the trajectories in phase space to initial conditions.

\item \textbf{Continuous phase transitions}: If there is a finite discontinuity in the first derivative of the thermodynamic potential, then the phase transition is termed first-order. During such a transition, a system either absorbs or releases a fixed amount of latent heat (e.g. the freezing/melting of water/ice). If the first derivative is continuous but higher derivatives are discontinuous or infinite, then the phase transition is called continuous, of the second kind, or critical. Examples include the critical point of the liquid-gas transition, the Curie point of the ferromagnetic transition, or the superfluid transition \citep{LandauLifshitz1980, Yeomans1992}. 

\item \textbf{Critical exponents}: Near the critical point, various thermodynamic quantities diverge as power laws with associated critical exponents. In equilibrium systems, there are scaling relations that connect some of the critical exponents of different thermodynamic quantities \citep{Cardy1996, LandauLifshitz1980, Sornette2004, Stanley1999,Yeomans1992}. 

\item \textbf{Critical phenomena}: Phenomena observed in systems that undergo a continuous phase transition. They are characterized by scale invariance: the statistical properties of a system on one scale are related to those at another scale only through the ratio of the two scales and not through anyone of the two scales individually.  The scale invariance is a result of fluctuations and correlations at all scales, which 
prevents the system from being separable in the large scale limit at the critical point  \citep{Cardy1996, Sornette2004, Yeomans1992}. 

\item \textbf{Declustering}: In studies of seismicity, declustering traditionally refers to the deterministic identification or fore-, main- and aftershocks in sequences (or clusters) of earthquakes clustered in time and space. Recent, more sophisticated techniques, e.g. stochastic declustering, assign  to earthquakes probabilities  of being triggered or spontaneous. 

\item \textbf{Dynamical scaling and exponents}: Non-equilibrium critical phase transitions are also characterized by scale invariance, scaling functions and critical exponents.  Furthermore, some evidence supports the claim that universality classes also exist for non-equilibrium phase transitions (e.g. the directed percolation and the Manna universality class in sandpile models), although a complete classification of classes is lacking and may in fact not exist at all. Much interest has recently focused on directed percolation, which, as the most common universality class of absorbing state phase transitions, is expected to occur in many physical, chemical and biological systems \citep{Hinrichsen2000, Luebeck2004, Sornette2004}.

\item \textbf{Finite size scaling}: If a thermodynamic or other quantity is investigated at the critical point under a change of the system size, the scaling behavior of the quantity with respect to the system size is known as finite size scaling \citep{Cardy1996}. The quantity may refer to a thermodynamic quantity such as the free energy or it may refer to an entire probability distribution function. At criticality, the sole length scale in a finite system is the upper cut-off $s_c$, which diverges in the thermodynamic limit $L \to \infty$. Assuming a lower cut-off $s_0 \ll s_c,s$, a finite size scaling ansatz for the distribution $P(s;s_c)$ of the observable variable $s$, which depends on the upper cut-off $s_c$ is then given by:
\be
P(s;s_c)=a s^{-\tau} G(s/s_c) \hspace{1cm} \text{for} ~~~s, s_c \gg s_0
\label{fss}
\ee
where the parameter $a$ is a non-universal metric factor, $\tau$ is a universal (critical) exponent, and $G$ is a universal scaling function that decays sufficiently fast for $s \gg s_c$ \citep{Cardy1996, Christensen-et-al2007}. \citet{Pruessner2004} provides a simple yet instructive and concise introduction to scaling theory and how to find associated exponents. System-specific corrections appear to sub-leading order. 

\item \textbf{Fractal}: A deterministic or stochastic mathematical object that is defined by its exact or statistical self-similarity at all scales. Informally, it often refers to a rough or fragmented geometrical shape which can be subdivided into parts which look approximately the same as the original shape. A fractal is too irregular to be described by Euclidean geometry and has a fractal dimension that is larger than its topological dimension but less than the dimension of the space it occupies. 

\item \textbf{Mean-Field}: An effective or average interaction field designed to approximately replace the interactions from many bodies by one effective interaction which is constant in time and space, neglecting fluctuations.

\item \textbf{Mechanisms for power laws}: Power laws may be the hallmark of critical phenomena, but there are a host of other mechanisms that can lead to power laws (see Chapter 14 of \citet{Sornette2004} for a list of power law mechanisms as well as \citet{Mitzenmacher2004,Clausetetal07}). Observations of scale invariant statistics therefore do not necessarily imply SOC, of course. Power laws express the existence of a symmetry (scale invariance) and there are many mechanisms by which a symmetry can be obtained or restored.

\item \textbf{Non-equilibrium phase transitions}: In contrast to systems at equilibrium, non-equilibrium phase transitions involve dynamics, energy input and dissipation. Detailed balance is violated and no known equivalent of the partition function exists,  from which all thermodynamic quantities of interest derive in equilibrium. Examples of non-equilibrium phase transitions include absorbing state phase transitions, reaction-diffusion models, morphological transitions of growing surfaces, and percolation in porous media  \citep{Hinrichsen2000, Luebeck2004}. 

\item \textbf{Phase transitions}: In (equilibrium) statistical mechanics, a phase transition occurs when there is a singularity in the free energy or one of its derivatives. Examples include the freezing of water, the transition from ferromagnetic to paramagnetic behavior in magnets, and the transition from a normal conductor to a superconductor \citep{LandauLifshitz1980,Yeomans1992}. 

\item \textbf{Renormalization group theory}: A mathematical theory built on the idea that the critical point can be mapped onto a fixed point of a suitably chosen transformation on the system's Hamiltonian. It provides a foundation for understanding scaling and universality and provides tools for calculating exponents and scaling functions. Renormalization group theory provides the basis for our understanding of critical phenomena \citep{Cardy1996,Stanley1999, Yeomans1992}.  It has been extended to non-Hamiltonian systems and provides a general framework for constructing theories of the macro-world from the microscopic description.

\item \textbf{Self-Organized Criticality (SOC)}: Despite two decades of research since its inception by \citet{Bak-et-al1987}  and the ambitious claim by \citet{Bak1996} that, as a mechanism for the ubiquitous power laws in Nature, SOC was "How Nature Works", a commonly accepted definition along with necessary and sufficient conditions for SOC is still lacking \citep{Jensen1998, Pruessner2004,Sornette2004}. A less rigorous definition may be the following: Self-organized criticality refers to a non-equilibrium, critical and marginally stable steady-state, which is attained spontaneously and without (explicit) tuning of parameters. It is characterized by power law event distributions and fractal geometry (in some cases) and may be expected in slowly driven, interaction-dominated threshold systems  \citep{Jensen1998}. Some authors additionally require that temporal and/or spatial correlations decay algebraically (e.g. \citep{Hergarten2002}, but see \citet{Pruessner2004}). Definitions in the literature range from broad (simply the absence of characteristic length scales in non-equilibrium systems) to narrow (the criticality is due to an underlying continuous phase transition with all of its expected properties) (see, e.g., \citet{PetersNeelin2006} for evidence that precipitation is an instance of the latter definition of SOC in which a non-linear feedback of the order parameter on the control parameter turns a critical phase transition into a self-organized one attracting the dynamics \citep{Sornette1992a}). 

\item \textbf{Spinodal Decomposition}: In contrast to the slow process of phase separation via nucleation and slow growth of a new phase in a material inside the unstable region near a first-order phase transition, spinodal decomposition is a non-equilibrium, rapid and critical-like dynamical process of phase separation that occurs quickly and throughout the material. It needs to be induced by rapidly quenching the material to reach a sub-area (sometimes a line) of the unstable region of the phase diagram which is characterized by a negative derivative of the free energy.

\item \textbf{Statistical physics} is the set of concepts and mathematical techniques allowing
one to derive the large-scale laws of a physical system from the specification of the relevant microscopic
elements and of their interactions.

\item \textbf{Turbulence}: In fluid mechanics, turbulence refers to a regime in which the dynamics
of the flow involves many interacting degrees of freedoms, and is very complex with intermittent
velocity bursts leading to anomalous scaling laws describing the energy transfer from injection at large scales to dissipation at small scales.


\item \textbf{Universality}: In systems with little or no frozen disorder, equilibrium continuous phase transitions fall into a small set of universality classes that are characterized by the same critical exponents and certain scaling functions become identical near the critical point. The class depends only on the dimension of the space and the dimension of the order parameter. For instance, the critical point of the liquid-gas transition falls into the same universality class as  the 3D Ising model. Even some phase transitions occurring in high-energy physics are expected to belong to the Ising class. Universality justifies the development and study of extremely simplified models (caricatures) of Nature, since the behavior of the system at the critical point can nevertheless be captured (in some cases exactly). However, non-universal features remain even at the critical point but are less important, e.g. amplitudes of fluctuations or system-specific corrections to scaling that appear at sub-leading order \citep{Cardy1996, Stanley1999,Yeomans1992, Zee2003}. 

\end{itemize}

\section{Definition and Importance of the Subject}

A fundamental challenge in many scientific disciplines concerns upscaling, that is,
of determining the regularities and laws of evolution at some large scale from those
known at a lower scale: biology (from molecules to cells, from cells to organs); 
neurobiology (from neurons to brain function), psychology (from brain to emotions, from evolution
to understanding), ecology (from species to the global web of
ecological interactions), condensed matter physics (from atoms
and molecules to organized phases such as solid, liquid, gas, and intermediate structures),
social sciences (from individual humans to social groups and to society),
economics (from producers and consumers to the whole economy), finance (from investors
to the global financial markets), Internet (from e-pages to the world wide web 2.0), semantics (from letters and words to sentences and meaning), and so on. Earthquake physics is no exception, with the challenge of 
understanding the transition from the laboratory scale (or even the microscopic and atomic scale)  
to the scale of fault networks and large earthquakes.

In this context, statistical physics has had a remarkably successful track record in addressing the
upscaling problem in physics. While the macroscopic laws of thermodynamics have been 
established in the 19th century, their microscopic underpinning were elaborated in the early 20th
century by Boltzmann and followers, building the magnificent edifice of statistical physics.
Statistical physics can  be defined as the set of concepts and mathematical techniques allowing
one to derive the large-scale laws of a physical system from the specification of the relevant microscopic
elements and of their interactions. Dealing with huge ensembles of elements (atoms, molecules)
of the order of the Avogadro number ($\simeq 6 \cdot 10^{23}$), statistical physics uses the
mathematical tools of probability theory combined with other relevant fields of physics to 
calculate the macroscopic properties of large populations.

One of the greatest achievement of statistical physics was the development of the 
renormalization group analysis, to construct a theory of interacting fields and of 
critical phase transitions. The renormalization group is a perfect example of how statistical
physics addresses the micro-macro upscaling problem. It decomposes a problem of finding the
macroscopic behavior of a large number of interacting parts into a succession
of simpler problems with a decreasing number of interacting parts,
whose effective properties vary with the scale of observation. The renormalization
group thus follows the proverb ``divide to conquer'' by organizing the
description of a system scale-by-scale. It is particularly adapted to critical
phenomena and to systems close to being scale-invariant. The renormalization
group translates into mathematical language the concept that the overall
behavior of a system is the aggregation of an ensemble of arbitrarily defined
sub-systems, with each sub-system defined by the aggregation of sub-subsystems,
and so on \citep{Sornette2004}.

It is important to stress that the term ``statistical'' in ``statistical physics'' has a very different
meaning, up to now, than in ``statistical seismology,'' a field that has developed as
a marriage between probability theory, statistics and the part of seismology concerned
with empirical patterns of earthquake occurrences \citep{Vere-Jones06} (but not with physics).
Statistical seismology uses stochastic models of seismicity, which 
are already effective large-scale representation of the dynamical organization.
In contrast, a statistical physics approach to earthquake strives to derive these
statistical models or other descriptions from the knowledge of the microscopic laws
of friction, damage, rupture, rock-water interations, mechano-chemistry and so on, 
at the microscopic scales \citep{Sornettereview,Sornettemechano}. In other words,
what is often missing in statistical seismology is the physics to underpin the stochastic model
on physically-based laws,  e.g. rate-and-state friction \citep{Diete94}.

The previously mentioned successes of statistical physics promote the hope that a similar
program can be developed for other fields, including seismology. The successes have been
more limited, due to the much more complex interplay between mechanisms, interactions and 
scales found in these out-of-equilibrium systems. This short essay provides a subjective entry to understand some of the different attempts, underlining the few successes, the problems and open questions. Rather than providing an exhaustive review, we mention what we believe to be important topics and have especially included recent work.

\section{Introduction}

Much of the recent interest of the statistical physics community has focused on applying scaling techniques, which are common tools in the study of critical phenomena, to the statistics of inter-event recurrence times or waiting times \citep{Bak-et-al2002, Corral2003, Corral2004a, Corral2004b, Corral2005a, Corral2005b, CorralChristensen2006, DavidsenGoltz2004, Livina-et-al2006}. However, the debate over  the relevance of critical phenomena to earthquakes stretches back as far as 30 years \citep{VereJones1977, Allegre-et-al1982, Smalley-et-al1985, Kagan1989, SornetteSornette1989, BakTang1989, SornetteSornette1990, Kagan1992, Olami-et-al1992, Kagan1994, SornetteSammis1995, Saleur-et-al1996a, Bak1996, Fisher-et-al1997, Dahmen-et-al1998, Jensen1998, NatureDebates1999, Hergarten2002, Sornette2002, Sornette2004, Kagan2006a, DahmenBenZion2008, Zoeller-et-al2008}. The current debate on recurrence statistics is thus the latest tack in an evolving string of arguments with a long history. As discussed below, many of the claims made in the recent articles on recurrence statistics have either been challenged, refuted or explained by previously known facts about earthquake statistics \citep{Lindman-et-al2005, Lindman-et-al2006, Molchan2005, SaichevSornette2006b, SaichevSornette2007, WernerSor07}. As will be discussed below, this debate in the literature is important because of the potential consequences for understanding earthquakes, but it needs to be pursued with rigorous scientific arguments accessible to both the seismological and the statistical physics communities. 

The debate would almost certainly benefit significantly from testing hypotheses with simulations to establish null hypotheses and benchmarks: seismicity patterns are sufficiently stochastic and earthquake catalogs contain a sufficient amount of observational uncertainties so as to make inference difficult. It is often not straightforward to predict the signal of well-known statistical features such as clustering in new data analysis techniques. Therefore, testing the purported claims by realistic simulations of earthquake catalogs can provide a strong benchmark against which the claims can be evaluated. This view and the
corresponding criticism of many studies has been put forward and defended for a long time by \citet{Kagan1999a}. 

Such a model-dependent approach may be at odds with the philosophy of a so-called ``model-free" analysis, which the community of statistical physicists claim to take in their analysis. For instance, network theory-based approaches, space-time window-based finite size scaling, box-covering methods and other techniques used in the study of critical and fractal phenomena are said to be ``model-free" because no assumptions about seismicity are supposedly made at the outset. By using model-free analysis techniques, the often uncertain and sometimes clearly wrong assumptions of flawed models and resulting biased results are meant to be circumvented. 

However, as is almost always the case in statistical hypothesis testing, the less assumptions are made about the test, the less powerful the test statistic. More importantly, seismicity is sufficiently stochastic so that well-known features may appear as novel in new analysis methods. Furthermore, to convince the seismological community of new data analysis techniques, the methods need to be tested on established knowledge and show the improvement over traditional methods. These types of initial tests are rarely performed by the statistical physics community. 

In the next section IV, we present  a summary of some of the concepts and calculational tools that have been developed
in attempts to apply statistical physics approaches to seismology. Then, section V summarizes the leading theoretical physical models of the space-time organization of earthquakes. Section VI  presents
a general discussion and several examples of the new metrics proposed by statistical physicists, underlining
their strengths and weaknesses.  Section VII briefly outlines future directions.

\section{Concepts and Calculational Tools}

\subsection{Renormalization, Scaling and the Role of Small Earthquakes in Models of Triggered Seismicity}

A common theme in many of the empirical relations in seismology (and in those employed in seismicity models) is the lack of a dominating scale. Many natural phenomena can be approached by the traditional reductionist approach to isolate a process at a particular scale. For example, the waves of an ocean can be described quite accurately by a theory that entirely ignores the fact that the liquid is made out of individual molecules. Indeed, the success of most practical theories in physics depends on isolating a scale \citep{Wilson1979}, although since this recognition, much progress has been made in developing a holistic approach for processes that do not fall into this class. Given current observational evidence, earthquakes seem to belong to the set of processes characterized by a lack of one dominating length scale: fluctuations of many or perhaps a wide continuum of sizes seem to be important and are in no way diminished -- even when interested solely in large-scale descriptions \citep{SP2003}. 

The traditional reductionist approach in seismology, which, for instance, attempted to separate large (main) shocks from small (fore- or after-) shocks, is slowly giving way to the holistic approach, in which all earthquakes are created equal and seismicity is characterized by fluctuations of all sizes. This gradual shift is supported, on a conceptual and qualitative level, by the vision of critical phenomena. A particularly strong model of the interactions between earthquakes has emerged in the concept of triggering, which places all earthquakes on the same footing: each earthquake can trigger its own events, which in turn can trigger their own events, and so on, according to the same probability distributions, and the resulting seismicity can be viewed as the superposed cascades of triggered earthquakes that cluster in space and time \citep{KaganKnopoff1981, Ogata1988, Ogata1998, HelmstetterSornette2002}. 

From this point of view, it is natural that small earthquakes are important to the overall spatio-temporal patterns of seismicity. Indeed, the scaling of aftershock productivity with mainshock magnitude suggests that small earthquakes are cumulatively as important for the triggered seismicity budget as rarer but larger events \citep{Felzer-et-al2002, Helmstetter2003, Helmstetter-et-al2005}. The importance of small earthquakes has also been documented in, e.g., \citep{Hanks1992, MichaelJones1998, Marsan2005}.

But earthquake catalogs do not contain information (by definition) about the smallest, unobserved events, which we know to exist from acoustic emission experiments and earthquakes recorded in mines. To guarantee a finite seismicity budget, \citet{SornetteWerner2005a} argued for the existence of a smallest triggering earthquake, akin to a ``ultra-violet cut-off" in quantum field theory, below which earthquakes do not trigger other events. Introducing a formalism which distinguishes between the detection threshold and the smallest triggering earthquake, they placed constraints on its size by using a simplified version of the popular Epidemic-Type Aftershock Sequence (ETAS) Model \citep{Ogata1988}, a powerful model of triggered seismicity based on empirical statistics, and by using observed aftershock sequences. \citet{SornetteWerner2005b} considered the branching structure of one complete cascade of triggered events, deriving an apparent branching ratio and the apparent number of untriggered events, which are observed when only the structure above the detection threshold is known. As a result of our inability to observe the entire branching structure, inferred clustering parameters are significantly biased and difficult to interpret in geophysical terms. Second, separating triggered from untriggered events, commonly known as declustering, also strongly depends on the threshold, so that it cannot even in theory constitute a physically sound method.  

\citet{SornetteWerner2005b} also found that a simplified, averaged version of the ETAS model can be renormalized onto itself, with effective clustering parameters, under a change of the threshold. \citet{SaichevSornette2006} confirmed these results for the stochastic number statistics of the model using a rigorous approach in terms of generating probability functions,  but also showed that the temporal statistics could not be renormalized. Furthermore, it can be shown (see Chapter 4 of \citet{Werner2007}) that the conditional intensity function of the ETAS model, the mathematical object which uniquely defines the model, cannot be renormalized onto itself under a change of magnitude threshold. It is not a fixed-point of the renormalization process operating via magnitude coarse-graining. The functional form of the model must change under a change in the detection threshold \citep{SaichevSornette2006}. In other words, if earthquakes occur according to an ETAS model above some cut-off $m_0$, then earthquakes above $m_d$ cannot be described by the ETAS model in a mathematically exact way. In practice, the ETAS model provides an excellent fit. The issue of how to deal with small earthquakes is thus reminiscent of the decades of efforts that have been invested in physics to deal with the famous
ultra-violet cut-off problem, eventually solved by the so-called ``renormalization''  theory of Feynmann, Schwinger and Tomonaga. In the 1960s and 1970s, this method of renormalization was extended into the ``renormalization group'' (in fact a semi-group in the strict mathematical sense) for the theory of critical phenomena (see glossary), which we also mention in section V. It is fair to say that there has been limited success in developing a multi-scale description of the physics of earthquakes and, in particular, in addressing the upscaling problem and the impact of the many small earthquakes 

One tantalizing approach, not yet really understood in terms of all its consequences and predictions,
is the variant of the ETAS model proposed by
\citet{VereJones2005}, which has the remarkable property of being bi-scale invariant under a transformation involving time and magnitudes. One of the modifications brought in by \citet{VereJones2005} is to assume that the distribution of the daughter magnitudes is dependent on the mother magnitude $m_i$ through a modification of the Gutenberg-Richter distribution 
of triggered earthquake magnitudes by a term of the form $\exp(-\delta |m-m_i|)$, where $\delta>0$
quantifies the distance to the standard Gutenberg-Richter distribution.
Remarkably, \citet{SaichevSornette2005b}, who studied the Vere-Jones model, found that, due to the superposition of the many magnitude distributions of each earthquake in the cascades of
triggered events, the resulting distribution of magnitudes over a stationary catalog is a pure Gutenberg-Richter law. Thus, there might be hidden characteristic scales in the physics of triggering
that are not revealed by the standard observable one-point statistical distributions.
Simulation and parameter estimation algorithms for the Vere-Jones model are not yet available.  If and when these algorithms become available, the study of this bi-scale invariant branching model may be a strong alternative to the ETAS model, as this model is exactly scale invariant with no ultra-violet nor infra-red cut-offs.

Nevertheless, being empirically based, these stochastic point process models lack a genuine microscopic physical foundation. The underlying physics is not explicitly addressed and only captured effectively by empirical statistics, even at the smallest scales. The physical processes and their renormalization are missing in this approach. 

\subsection{Universality}

Universality, as defined in the glossary, justifies the development and study of extremely simplified models (caricatures) of Nature, since the behavior of a studied system at the critical point can nevertheless be captured by toy models (in some cases exactly). For instance, the liquid-gas transition, the ferromagnetic to paramagnetic transition, and the behavior of binary alloys, all apparently different systems, can be described successfully by an extremely simplified picture of Nature (the Ising model) because of universality. The hope that a similar principle holds for earthquakes (and other non-equilbrium systems)  underpins many of the models and tools inspired by statistical physics that have been applied to seismicity, to bring about the ``much coveted revolution beyond reductionism" \citep{Barabasi-et-al1999}. 

Speaking loosely, the appearance of power laws in many toy models is often interpreted as a kind of universal behavior. The strict interpretation of universality classes, however, requires that critical exponents along with scaling functions are identical for different systems. It is interesting to note that slight changes in the sand-pile model already induce new universality classes, so that even within a group of toy models, the promise of universality is not, strictly, fulfilled \citep{Bak-et-al1987, Kadanoff-et-al1989, Manna1991}. 

A lively debate in seismology concerns the universality of, on one hand, the frequency-size distribution of earthquake magnitudes (e.g. \citep{Kagan1993, Wesnousky1994, Fisher-et-al1997, Dahmen-et-al1998, Kagan1999a, JacksonKagan2006}, and, on the other hand, the universality of the exponent of the Gutenberg-Richter distribution (e.g. \citep{Sornette-et-al1991,BirdKagan2004,WiemerKatsumata1999, Schorlemmer-et-al2005}). A spatio-temporally varying critical exponent is not traditionally part of the standard critical phenomena repertoire, although analytical and numerical results based on a simple earthquake model on a fault showed a possible spontaneous switching between Gutenberg-Richter and characteristic earthquake behavior associated with a non-equilibrium phase transition \citep{Fisher-et-al1997, Dahmen-et-al1998, BenZion-et-al1999, DahmenBenZion2008, Zoeller-et-al2008}. 
Another possible mechanism for the coexistence of and intermittent shifts between different regimes (Gutenberg-Richter scaling and characteristic earthquakes) stems from 
the competition between several interacting faults which give rise
to long intervals of activity in some regions followed by similarly long intervals of quiescence
\citep{Sornette-et-al1994,Sornette-et-al1995,Leeetal99}.
Both careful empirical investigations of seismicity parameters and theoretical progress on heterogeneous, spatially extended critical phenomena may help elucidate the controversy.

\subsection{Intermittent Periodicity and Chaos}

As part of the conquest of chaos theory in the 70s and 80s \citep{May1976, Eckman1981}, its concepts and methods were invariably also applied to seismicity. \citet{HuangTurcotte1990, HuangTurcotte1990b} modeled the interaction between two faults by two sliding blocks that are driven by a plate through springs, coupled to each other via another spring and endowed with a velocity-weakening friction law. The dynamical evolution of the blocks showed chaotic behavior and period-doubling (the Feigenbaum route to chaos \citep{Eckman1981}). \citet{HuangTurcotte1990, HuangTurcotte1990b} suggested that the interaction of the Parkfield segment with the southern San Andreas fault may be governed by the kind of chaotic behavior they observed in their model: the lighter block slipped quasi-periodically for several times until both slipped together. Their study explained that apparent quasi-periodicity of earthquakes on fault segments may be a result of chaotic interactions between many fault segments, thereby providing a warning that the extrapolations of quasi-periodic models are not to be trusted (e.g. \citep{HarrisRamon2006, Bakun-et-al2005} and references therein). However, it is doubtful that models with just a few degrees of freedom can go a long way towards providing
deeper physical insights, or predictive tools. One needs to turn to models with a large number of degrees of freedom, for which turbulence appears as the leading paradigm of complexity.

\subsection{Turbulence}

A drastically different approach has been favored by Yan Kagan \citep{Kagan1989, Kagan1992, Kagan1994, Kagan2006a}, who described seismicity as the ``turbulence of solids" - attesting to the potentially far more complex problems that need to be solved than the theory of critical phenomena promises to deliver. While renormalization group methods and scaling theory have contributed immensely to the study of turbulence \citep{Frisch1995}, the problem of turbulence involves significant additional complications. First, loosely speaking, renormalization group theory  helps predict global behavior by coarse-graining over degrees of freedom, which is essentially a bottom-up approach. In turbulence, the enstrophy acts bottom-up, but the energy cascades top to bottom. Secondly, there is a significant spatial and topological aspect to turbulence, for instance involving topological defects, such as filament structures, which are crucial to the dynamical evolution of the system. 
The existence of the two cascades (top-down and bottom-up) as well as the influence of the dissipation
scale all the way within the so-called inertial range makes turbulence the most important classical problem
still unsolved in Physics. The importance of addressing the issue of the interplay between the
top-down and bottom-up cascades in earthquake toy models has been outlined by \citet{Gab1,Gab2,Zal1,Zal2}.

The analogous problem for seismicity lies in the complex fault network, which constrains seismicity through its weak structures but also grows and evolves because of earthquakes redistributing stresses and rupturing fresh surfaces. The statistical description of this tensorial and dynamical problem is only at its beginning \citep{KaganKnopoff1980, Kagan1981a, Kagan1981b, KaganKnopoff1985a, KaganKnopoff1985b, Kagan1987, Kagan1987b, Kagan1988, Kagan1992b, Miltenberger-et-al1993, Sornette-et-al1994, Gabrielov-et-al1996}. But it is likely to be a key aspect to the dynamical evolution of faults and seismicity. New physics
and approaches are required to tackle the tensorial nature of the stress and strain fields and the complex topological structures of defects, from dislocations to joints and faults, as well as the many different physical processes operating from the atomic scale to the macro-scale \citep{Sornettereview,Sornettemechano}.

\subsection{Self-Organized Criticality}

Self-organized criticality (SOC) refers to
the spontaneous organization of a system driven from outside into a globally dynamical
statistical stationary state, which is characterized by self-similar
distributions of event sizes and fractal geometrical properties.
SOC applies to the class of phenomena occurring in 
driven out-of-equilibrium systems made of many interactive components, which
possess the following fundamental properties : 1) a highly non-linear behavior,
2) a very slow driving rate , 3) a globally stationary  regime, 
characterized by stationary statistical properties,
and 4) power-law distributions of event sizes and fractal geometrical properties.
The crust obeys these four conditions, as first suggested by
\citet{SornetteSornette1989} and \citet{BakTang1989}, who proposed to understand
the spatio-temporal complexity of earthquakes in this perspective.

The appeal of placing the study of earthquakes in the framework of critical phenomena may be summarized as follows. Power law distributions can be understood as a result of an underlying continuous phase transition into which the crust has organized itself \citep{SorDavySor,SornetteGeilo}. Applying the methods of renormalization group theory may help calculate exponents and scaling functions and rationalize the spatio-temporal organization of seismicity along with its highly correlated structures. For instance, \citet{SorVirieux} provided a theoretical framework which links the scaling  laws exhibited by earthquakes at short times and plate tectonic deformations at large times. Perhaps earthquakes fall into a universality class which can be solved exactly and/or investigated in toy models. Moreover, studying the detailed and highly complicated microphysics involved in earthquakes may not lead to insights about the spatio-temporal organization, because, as a critical phenomenon, the traditional approach of separating length scales to describe systems is inadequate. On the other hand, as mentioned above, there is the possibility
of a hierarchy of physical processes and scales which are inter-related \citep{Ouillonetal1995,Ouillonetal1996}, for which the
simplifying approach in terms of critical phenomena is likely to be insufficient.  

As another reason for the importance of the topic, interesting consequences for the predictability of earthquakes might be derived, for instance by mapping earthquakes to a genuine critical point (the accelerating moment release hypothesis,  e.g. \citep{SornetteSammis1995, Sornette2002}) or by mapping earthquakes to SOC (e.g. \citep{Geller-et-al1997a, NatureDebates1999}). The latter mapping had
led some to argue that earthquakes are inherently unpredictable. In the sandpile paradigm 
\citep{Bak-et-al1987}, there is little difference between small and large avalanches, and this led
similarly to the concept that ``large earthquakes are small earthquakes that did not stop,'' hence
their supposed lack of predictability.
More than ten years after this contentious proposal, a majority of researchers, including most
of the authors of this ``impossibility claim,'' recognize that there is some degree of
predictability \citep{Jordan2006,Helmstetter-et-al2006}. Actually, the clarifications came
from investigators of SOC, who recognized that the long-term organization of sandpiles
\citep{Dhar90,Dhar99}  and of toy models of earthquakes and fault networks \citep{Miltenberger-et-al1993,Sornette-et-al1994,Sornette-et-al1995}
is characterized by long-range spatial and temporal correlations.
Thus, large events may indeed be preceded by subtle long-range organizational structures, 
an idea at the basis of the accelerating moment release hypothesis. This idea is also
underlying the pattern recognition method introduced by \citet{Gelfandetal76} and 
developed extensively by V. Keilis-Borok and his collaborators for
earthquake predictions  \citep{KBetal2003}.
In addition, \citet{Huangetal98} showed that avalanche dynamics 
occurring within hierarchical geometric structures 
are characterized by significant precursory activity before large events; this provides a clear
proof of the possible coexistence between critical-like precursors of large events and a long-term
self-organized critical dynamical state.

In summary, self-organized criticality provides a general conceptual framework to articulate
the search for a physical understanding of the large-scale and long-time statistical
properties of the seismogenic process and of the predictability
of earthquakes. Beyond this, it is of little help as many different mechanisms have been
documented at the origin of SOC (see, e.g.,  chapter 15 in \citep{Sornette2004}. SOC is not
a theory, it does not provide any specific calculation tools; it is a concept offering
a broad classification of the kinds of dynamics that certain systems, including the Earth crust,
seem to spontaneously select.


\section{Competing mechanisms and models}

It should be noted at this point that the statistical physics approach to earthquake science is not limited to SOC. Over the years, several groups have proposed to apply the concepts and tools
of statistical physics to understanding the multiscale dynamics of earthquake and fault systems.
Various mechanisms drawn conceptually from statistical mechanics but not necessary even limited to critical (phase transition) phenomena have been proposed and are being pursued. Such approaches include the concept of the critical point earthquake related to accelerated moment release, network theory, percolation and fiber models as models for fracture, and many more, some of which can be found in \citep{Hergarten2002, Sornette2004, Turcotte1997, Turcotte-et-al2000}. 

In this section, we outline some of the major model classes which underpin
distinct views on what are the dominating mechanisms to understand earthquakes
and their space-time organization.

\subsection{Roots of complexity in seismicity: dynamics or heterogeneity?}

The 1990s were characterized by vigorous discussions 
at the frontier between seismology and statistical
physics aimed at understanding the origin of the observed complexity of the spatio-temporal
patterns of earthquakes. The debate was centered on the question of whether 
space-time complexity can occur on a single homogeneous fault, solely 
as a result of the nonlinear dynamics \citep{Shaw93,Shaw94,Shaw95,Shaw97,CochardMada94,CochardMada96,Langeretal96,BenzionRice97},
associated with the slip and velocity dependent friction law validated empirically in particular by
\citet{Diete87,Diete92,Diete94,DietKil94}. Or, is the presence
of quenched heterogeneity necessary \citep{Rice93,BenzionRice93,BenzionRice95,Knopoff1996}?

The rediscovery of the multi-slider-block-spring model of \citet{BurridgeKnopoff64}
by \citet{CarlsonLanger89} led to a flurry of investigations by physicists \citep{CarlsonLanger94,RundleKlein95,Langeretal96}, finding 
an enticing entry to this difficult field, in the hope of capturing the main empirical statistical properties
of seismicity.  It is now understood that complexity in the stress field, in co-seismic slips and in sequences
of earthquakes can emerge purely 
from the nonlinear laws. However, 
heterogeneity is probably the most important factor dominating the multi-scale
complex nature of earthquakes and faulting \citep{Ouillonetal1995,Ouillonetal1996,Scholz91,Scholz02}.
It is also known to 
control the appearance of self-organized critical behavior in a class of models
relevant to the crust \citep{Sornette-et-al1995,ShnirmanBlanter98}.

\subsection{Critical earthquakes}

This section gives a brief history of the ``critical earthquake'' concept.

We trace the ancestor of the critical earthquake concept to
\citet{VereJones1977}, who used a branching
model to illustrate that rupture can proceed through a cascade of damage events.
\citet{Allegre-et-al1982} proposed what is in essence a percolation model of
damage/rupture describing the state of the crust before an earthquake.  
They formulated the model using the language of 
real-space renormalization group, in order to 
emphasize the multi-scale nature of the underlying physics, and the incipient
rupture as the approach to a critical percolation point. Their approach
is actually a reformulation in the language of earthquakes of the real-space renormalization
group approach to percolation developed by \citet{Reynoldsetal77}. \citet{Chelidze82} 
independently developed similar ideas.
In the same spirit, \citet{Smalley-et-al1985} proposed a renormalization group treatment
of a multi-slider-block-spring model. \citet{SornetteSornette1990} took seriously the concept put
forward by \citet{Allegre-et-al1982} and proposed to test it empirically by searching
for the predicted critical precursors. \citet{Voight88,Voight89}
was probably the first author to introduce the idea of a time-to-failure analysis quantified by a second order
nonlinear ordinary differential equation.  For certain values of the
parameters, the solution of \citet{Voight88,Voight89}' s time-to-failure
equation takes the form of a finite time singularity (see \citet{SammisSornette02}
for a review and \citet{SorHelm} for a mechanism based on the
ETAS model).  He proposed and did use it later to predict volcanic
eruptions.  The concept that earthquakes are somehow associated with 
critical phenomena was also underlying the research efforts of a part of the russian school
\citep{KBed90,Tumarkin92}.

The empirical seed for the critical earthquake concept was the
repeated observations that large earthquakes have been preceded by an increase in the number of
intermediate size events 
\citep{KeilisMali64,Mogi69,Lindh90,Ellsworth81,Raleigh82,Keilis88,SykesJaume90,Jones94, BrehmBraille1998, JaumeSykes1999}. 
The relation between these intermediate events and the subsequent main event took a long time
to be recognized because the precursory events occur over such a large area.
\citet{SykesJaume90} proposed a
specific law $\sim \exp[t/\tau]$  quantifying the acceleration of seismicity prior to
large earthquakes. 
\citet{BufeVarnes93} proposed that the finite-time singularity law
\begin{equation}
\epsilon_{Benioff} \sim   1/(t_c-t)^m
\label{hkghkwgb}
\end{equation}
is a better empirical model than the exponential law.  In (\ref{hkghkwgb}),
$\epsilon_{Benioff}$
is the cumulative Benioff strain, $t_c$ is critical time of the occurrence of the target earthquake
and $m$ is a positive exponent. The fit  with this law of the empirical Benioff strain calculated by summing
the contribution of earthquakes in a given space-time window is supposed to provide the time $t_c$ of the earthquake and thus constitutes a prediction.
This expression (\ref{hkghkwgb}) was justified by a 
mechanical model of material damage. It is important to understand
that the law (\ref{hkghkwgb}) can emerge as a consequence of a variety
of mechanisms, as reviewed by  \citet{SammisSornette02}.

One of these mechanisms has been coined the ``critical earthquake'' concept, first
formulated by
\citet{SornetteSammis1995}, who proposed to reinterpret the formula (\ref{hkghkwgb})
proposed by \citet{BufeVarnes93} and 
previous related works by generalizing them within the statistical physics
framework.
This concept views a large earthquake
as a genuine critical point. Using the insight of critical points in
rupture phenomena, \citet{SornetteSammis1995} proposed to enrich equation (\ref{hkghkwgb}),
now interpreted as a kind of diverging susceptibility in the sense of critical phenomena,
by considering complex exponents (i.e. log-periodic corrections to
scaling). These structures accommodate the possible presence of a hierarchy of 
characteristic scales, coexisting with power laws expressing the scale invariance
associated with a critical phenomenon \citep{Sornette1998}.
This was followed by several extensions 
\citep{Saleur-et-al1996a,Huangetal98,Johansen1,Johansen2}.
\citet{Sornette2002} reviewed the concept of critical ``ruptures''  and earthquakes with application
to prediction. \citet{IdeSor02} presented a simple dynamical mechanism to obtain finite-time singularities (in rupture in particular) decorated by complex exponents (log-periodicity).   \citet{bowmanetal98,OuillonSor00,Zoeller-et-al2001, ZoellerHainzl2002,OuillonSor04} proposed empirical tests of the critical earthquake concept.
The early tests of \citet{bowmanetal98} have been criticized by \citet{Hardebeck07}, while \citet{VereJones-et-al2001} commented on the lack of a formal statistical basis of the accelerating moment release model. This stresses the need for rigorous tests in the spirit of 
\citep{Rhoades04,Rhoades05}. The debate is wide open, especially in view of the recent developments to improve the determination of the relevant spatio-temporal domain that should be used to perform
the analyses \citep{BowmanKing01a,BowmanKing01b,KingBowman03,Levinetal06} (see 
\citep{FreundSornette07} for a review).

\subsection{Spinodal decomposition}

Klein, Rundle and their collaborators have suggested a mean-field approach to the 
multi-slider-block-spring model justified by the long-range nature of the elastic interactions
between faults. This has led them to propose that the fluctuations of the strain and stress
field associated with earthquakes are technically those occurring close to a
spinodal line of an underlying first-order phase transition (see \citep{RundleKlein1993,Kleinetal97,Rundletal03} and references therein).
This conceptual view has inspired them to develop the  ``Pattern Informatics'' technique, an empirical seismicity forecasting method based on the 
idea that changes in the seismicity rate are proxies for changes in the underlying stress
\citep{Tiampetal06,Hollidayetal06}.

The fluctuations associated with a spinodal line are very similar to those observed
in critical phenomena. It is thus very difficult if not impossible in principle to falsify
this hypothesis against the critical earthquake hypothesis, since both are expected
to present similar if not identical signatures. Perhaps, the appeal of the spinodal
decomposition proposal has to be found at the theoretical level, from the fact that first-order phase transitions
are more generic and robust than critical phenomena, for systems where heterogeneity and
quenched randomness are not too large.

\subsection{Dynamics, stress interaction and thermal fluctuation effects}

\citet{Fisher-et-al1997, Dahmen-et-al1998, BenZion-et-al1999} and co-workers (see the reviews by \citet{DahmenBenZion2008} and \citet{Zoeller-et-al2008}) have introduced a mean-field
model (resulting from a uniform long-range Green function) of a single fault, whose
dynamical organization is controlled by
two control parameters, $\epsilon$ which measures the dynamic stress weakening 
and $c$ which is the deviation from stress conservation (due for instance to coupling with ductile layers).
The point $(\epsilon =0; c=0)$ is critical in the sense of a phase
transition in statistical physics, with its associated scale invariant fluctuations
described by power laws. Dynamic stress strengthening ($\epsilon <0$) leads
to truncated Gutenberg-Richter power laws. Dynamic stress weakening ($\epsilon >0$) 
is either associated with a truncated Gutenberg-Richter power law for $c>0$ or with characteristic
earthquakes decorating a truncated power law for $c<0$. The coexistence of 
a characteristic earthquake regime with a power-law regime is particularly interesting as it 
suggests that they are not exclusive properties but may characterize the same underlying
physics under slightly different conditions. This could provide a step towards 
explaining the variety of empirical observations in seismology \citep{Schwartz84,Wesnousky1994,Kagancharaearh94,Aki95}.

\citet{Sornette-et-al1995} have obtained similar conclusions using a quasi-static model
in which faults grow and self-organize into optimal structures by repeated earthquakes.
Depending on the value of dynamical stress drop (controlling the coupling strength between elements) relative to the amplitude of the frozen heterogeneity of the stress thresholds controlling the earthquake nucleation on each fault segment,  a characteristic earthquake 
regime with a truncated power law is found for small heterogeneity or large stress drop while the power law SOC regime is recovered by large heterogeneity or small stress drop. The two approaches of
\citep{Fisher-et-al1997, Dahmen-et-al1998, BenZion-et-al1999, DahmenBenZion2008} on the one hand and of \citet{Sornette-et-al1995} on the other hand can be reconciled  conceptually by noting
that the dynamic stress weakening of \citet{DahmenBenZion2008} controls the dynamical generated
stress heterogeneity while the lack of stress conservation $c$ controls the coupling strength.  
Fundamentally, the relevant control parameter is the degree of coupling between fault elements
seen as threshold oscillators of relaxation versus the variance of the disorder in their
spontaneous large earthquake recurrence times. Generically, power
law statistics are expected to co-exist with synchronized behavior in a general
phase diagram in the heterogeneity-coupling strength plane \citep{Sornette-et-al1995,Osorio07}.

Let us finally mention a promising but challenging theoretical approach, which has the ambition
to bridge the small-scale physics controlled by thermal nucleation of rupture to
the large-scale organization of earthquakes and faults \citep{OuiSor05,SorOui05}. Partial
success has been obtained with a remarkable prediction on the (``multifractal'') dependence of the Omori law 
exponent of aftershocks on the magnitude of the mainshock, verified by 
careful empirical analyses on earthquakes in California, Japan and worldwide \citep{OuiRibeiroSor07}.


\section{Empirical studies of seismicity inspired by statistical physics}

{\it \small ``False facts are highly injurious to the progress of science, for they often
endure long; but false views, if supported by some evidence, do little harm,
for everyone takes a salutary pleasure in proving their falseness.''
Charles Darwin, in The Origin of Man, Ch. 6.}
\vskip 0.3cm

\subsection{Early successes and latter subsequent challenges}

A significant benefit of the statistical physics approach to seismology has been the introduction
of novel techniques to analyze the available empirical data sets, with the goal of obtaining
new insights into the spatio-temporal organization of seismicity and of revealing novel regularities and laws
that may guide the theoretical analysis. 

A prominent forerunner is the application of 
the concept of fractals introduced by \citet{Mandelbrot82} and of the measures of fractal dimensions
to describe complex sets of earthquake epicenters, hypocenters and fault patterns. The use of fractals
has constituted an epistemologic breakthrough in many fields, and not only in seismology.
Indeed, before Mandelbrot, when dealing with most complex
systems, one used to say: ``this is too complicated for a quantitative analysis''
and only qualitative descriptions were offered. After Mandelbrot, one could hear: ``this is a fractal, 
with a fractal dimension equal to xxx!'' By providing a new geometrical way of thinking about
complex systems associated with novel metrics, Mandelbrot and his fractals have
extended considerably the reach of quantitative science to many complex systems in all fields. 

However, while there have been some attempts to use fractal dimensions as 
guidelines to infer the underlying organization processes, as for instance in \citep{SornetteA1,SornetteA2}, 
most of the initial reports have lost their early appeal \citep{Turcottefractal86,ScholzMandel89,BartonLaPointe1,BartonLaPointe2} since the complexity of seismicity and faulting is much too great to be captured by scaling laws embodying solely a simple scale invariance symmetry. Among others, multifractal and adapted wavelet tools are needed to quantify this complexity, see for instance \citep{Geilikman90,Ouillonetal1995,Ouillonetal1996, MolchanKonrod2005, OuillonDucorSor08}. It should also be noted that few studies of the fractal dimensions of seismicity address the significant issues of errors, biases and incomplete records in earthquake catalogs - a notable exception being \citep{Kagan2007b}.

Since the beginning of the 21st century, a renewal of interest and efforts have burgeoned as groups of statistical physicists, interested in earthquakes as a potential instance of self-organized criticality (SOC), have claimed ``novel", ``universal" and ``robust" scaling laws from their analysis of the spatio-temporal organization of seismicity. The authors purport to have discovered universal and hitherto unknown features of earthquakes that give new insights into the dynamics of earthquakes and add to the evidence that earthquakes are self-organized critical. We now discuss a few of these recent studies to illustrate the existence of potential problems in the ``statistical physics'' approach. In a nutshell, we show that perhaps most of these ``novel scaling laws" can be explained entirely by already known statistical seismicity laws. This claim has been defended by other experts of statistical seismology, the most vocal being perhaps Yan Kagan at UCLA \citep{Kagan1999a}.

The flurry of interest from physicists comes from their fascination with the self-similar properties exhibited by seismicity (e.g. the Gutenberg-Richter power law of earthquake seismic moments, the Omori-Utsu law of the decay of aftershock rates after large earthquakes, the fractal and multifractal space-time organization of earthquakes and faults, etc.), together with the development of novel concepts and techniques that may provide new insights. But, and this is our main criticism based on several detailed examples discussed below, many of the new approaches and results do not stand close scrutiny. This failure is rooted in two short-comings: (i) the lack of testing of new methods on synthetic catalogs generated by benchmark models which are based on well-known statistical laws of seismicity, and (ii) the failure to consider earthquake catalog bias, incompleteness and errors. The latter may cause catalog artifacts to appear as genuine characteristics of earthquakes. Testing the results on a variety of catalogs and considering the influence of various catalog errors can help minimize their influence. The former short-coming often leads to the following scenario: authors fail to realize that a simpler null hypothesis could not be rejected, namely that their ``discovery'' could actually be explained by just a combination of  basic statistical laws known to seismologists for decades.

The well-established laws of statistical seismicity that authors should consider before claiming for novelty include the following:
\begin{enumerate}
\item the Gutenberg-Richter law for the distribution of earthquake magnitudes with a $b$-value close to $1$ (corresponding to an exponent $\simeq 2/3$ for the probability density function of seismic moments), 
\item the Omori-Utsu law for the decay of the rate of aftershocks following a mainshock,
\item the inverse Omori law for foreshocks,
\item the fact that aftershocks also trigger their own aftershocks and so on, and that aftershocks do not
seem to exhibit any distinguishable physical properties,
\item the fact that the distribution of distances between mainshocks and aftershocks has a power law tail, 
\item the fertility law (the fact that earthquakes of magnitude $M$ trigger of the order of $10^{aM}$ aftershocks with $a \leq b \simeq 1$, 
\item the fractal distribution of faults which are concentration centers for earthquakes. 
\end{enumerate}
This above non-exhaustive list selects ``laws'' which are arguably non-redundant, in the sense
that it is likely not possible to derive one of these laws from the others (a possible exception is
the inverse Omori law for foreshock, which can be derived from the direct Omori law
for aftershocks in the context of the ETAS model \citep{HelmSorGras03,HelmSor03a}).
Some experts would argue that we should add to this list other claimed regularities, such as ``B{\aa}th's law'' (see e.g. \citep{ShcherbakovTurcotte04} for a recent discussion emphasizing the importance
of this law), that states that the differences in magnitudes between mainshocks and their largest aftershocks are approximately constant, independent of the magnitudes of mainshocks \citep{Baath1965}. However,  \citet{HelmSor03b,SaiSorBath05} have shown that Bath's law can be accurately recovered
in ETAS-type models combining the first, second, fourth, and sixth laws stated above, with the assumption that any earthquake can trigger subsequent earthquakes.

\subsection{Entropy method for the distribution of time intervals between mainshocks}

\citet{Megaetal03} used the ``diffusion entropy'' method to argue for a power-law distribution of time intervals between a large earthquake (the mainshock of a seismic sequence or cluster) and the next one.
\citet{helmsormega04}  showed that all the ``new" discoveries reported by \citet{Megaetal03} (including the supposedly new scaling) can be explained solely by Omori's law for intra-cluster times, without correlation between clusters, thus debunking the claim for novelty.

\subsection{Scaling of the PDF of Waiting Times}

\citet{Bak-et-al2002} analyzed the scaling of the probability density function of waiting times between successive earthquakes in southern California as a function of ``box size" or small regions in which subsequent earthquakes are considered. They found an approximate collapse of the pdfs for different seismic moment thresholds $S$ and box sizes $L$ which suggested the following scaling ansatz for the waiting times $T$: 
\be
T^{\alpha} P_{S,L}(T)=f(T S^{-b} L^{d_f})
\label{bakscaling}
\ee
where $b = 1$ is the Gutenberg-Richter exponent, $d_f \simeq 1.2$ was claimed to be a spatial fractal dimension of seismicity (see \citet{MolchanKonrod2005} and \citet{Kagan2007b} for more in-depth studies), $\alpha = 1$ was identified as the exponent in the Omori law and $f(\cdot)$ is a scaling function which was proposed to be roughly constant up to a constant (``kink") beyond which it quickly decays. The scaling (\ref{bakscaling}) was claimed to be a unified law for earthquakes that revealed a novel feature in the spatio-temporal organization of seismicity in that the Gutenberg-Richter, the Omori law and the spatial distribution of earthquakes were unified into a single picture that made no distinction between fore-, main- and aftershocks. The scaling relations and critical exponents were claimed to be contained in the scaling ansatz. \citet{Corral2003, Corral2004a, Corral2004b, Corral2005a} and others broadened the analysis to other regions of the world. \citet{Corral2004a} proposed a slightly different scaling ansatz for a modified data analysis. 

Early criticism came from \citet{Lindman-et-al2005}, who noted that generating synthetic data using a non-homogeneous Poisson process derived from Omori's law was able to reproduce some of the results of \citet{Bak-et-al2002}, indicating a rather trivial origin of the unified scaling law. \citet{Molchan2005} showed that, if at least two regions in the data set are independent, then, if a scaling relation were to hold exactly, this scaling function could only be exponential. All other functions could only result in approximate data collapses. Proponents of the unified scaling law, e.g. \citet{Corral2005b},  argued that indeed all regions were correlated, as expected in systems near a critical point  so that the assumption of independence between different regions should not hold. But \citet{Molchan2005} also showed that a simple Poisson cluster model (Poissonian mainshocks that trigger Omori-type aftershock sequences) could reproduce the short and long time limits of the observed statistics, indicating that the Omori law, the Gutenberg-Richter relationship and simple clustering were the sole ingredients necessary for the observed short and long time limit, and no spatial correlation was needed.  

\citet{SaichevSornette2006b, SaichevSornette2007} extended Molchan's arguments to show that the approximate data collapse of the waiting times could be explained completely by the Epidemic-Type Aftershock Sequence (ETAS) model of \citet{Ogata1988}. This provided further evidence that the apparent data collapse was only approximate. Remarkably, the theoretical predictions of the ETAS model seem to fit the observed data better than the phenomenological scaling function proposed by \citet{Corral2004a} to fit the data. \citet{SaichevSornette2006b, SaichevSornette2007} thus showed that a benchmark model of seismicity was able to reproduce the apparent unified scaling law and that therefore the distribution of interevent times did not reveal new information beyond what was already known via statistical laws: The combination of the Gutenberg-Richter law, the Omori law, and the concept of clustering suffice to explain the apparent ``universal" scaling of the waiting times. 

\citet{SorUtkinSai08} developed an efficient numerical scheme to solve accurately the set of nonlinear integral equations derived previously in \citep{SaichevSornette2007} and found a dramatic lack of power for the distribution of inter-event times to distinguish between quite different sets of parameters, casting doubt on the usefulness of this statistics for the specific purpose of identifying the clustering parameter (e.g. \citep{Hainzl-et-al2006b}).

\subsection{Scaling of the PDF of Distances Between Subsequent Earthquakes}

\citet{DavidsenPaczuski2005}  claimed evidence contradicting the theory of aftershock zone scaling in
favor of scale-free statistics. Aftershock zone scaling refers to the scaling of the mainshock rupture length, along which most aftershocks occur, with the mainshock magnitude \citep{Kagan2002a}. \citet{DavidsenPaczuski2005} suggested that the probability density function of spatial distances between successive earthquakes obeys finite size scaling with a novel dynamical scaling exponent, suggesting that the mainshock rupture length scale has no impact on the spatial distribution of aftershocks and that earthquakes are self-organized critical.

\citet{WernerSor07} debunked this claim by showing that (i)  the purported power law
scaling function is not universal as it breaks down in other regions of the world; (ii) the results obtained
by \citet{DavidsenPaczuski2005} for southern California depend crucially on a single earthquake (the June 28, 1992, M7.3 Landers earthquake): without Landers and its aftershocks, the power law
disappears; (iii) a model of clustered seismicity, with aftershock zone scaling explicitly
built in, is able reproduce the apparent power law, indicating that an apparent lack
of scales in the data does not necessarily contradict aftershock zone scaling and the
existence of scales associated with mainshock rupture length scales.

\subsection{The Network Approach}

The recent boom in the statistical mechanics of network analysis has recently extended to applications well beyond physics (for reviews, see  \citep{BarabasiAlbert1999, AlbertBarabasi2002, DorogevtsevMendes2003, Newman2003}). Earthquake seismology is no exception \citep{BaiesiPaczuski2004, PeixotoPrado2004, AbeSuzuki2004, AbeSuzuki2004b, BaiesiPaczuski2005, AbeSuzuki2005, Baiesi2006, AbeSuzuki2006,PeixotoPrado2006}. The resulting impact has been limited so far for several reasons. 

A major concern is the assumption that earthquake catalogs as downloaded from the web are data sets fit for immediate analysis. \citet{Kagan2003,Helmstetter-et-al2005,Werner2007} and \citet{WernerSor08} present modern and complementary assessments of the many issues of incompleteness spoiling even the best catalogs. In particular, we should stress that magnitude determinations are surprisingly inaccurate, leading to large errors in seismic rate estimates \citep{WernerSor08}. Furthermore, there is no such thing as a complete catalog above a so-called magnitude of completeness, due to the fact that a non-negligible fraction of earthquakes are missed 
in the aftermath of previous earthquakes \citep{Kagan2003,Helmstetter-et-al2005}. One should be
concerned that analyses in terms of network metrics could be particularly sensitive to these defects. Nevertheless,  \citep{AbeSuzuki2004, AbeSuzuki2004b, AbeSuzuki2005, AbeSuzuki2006} applied metrics of network analysis to ``raw" catalogs which included events well below the estimated magnitude of completeness. As a result of neglecting to use a (reasonably) homogeneous and trustworthy data set, the results of their analysis may be severely biased, because the reliability of the inferred network structure is likely more sensitive than other metrics to the correct spatio-temporal ordering of the earthquake catalog. No serious study has yet been performed to quantify the usually serious impact of quality issues on the metrics used in network analysis. As a consequence, it is also not clear how to interpret the ``success" of \citep{PeixotoPrado2004, PeixotoPrado2006} in reproducing the ``features" of Abe and Suzuki's analysis on the synthetic seismicity generated by a spring-block model.

 
In addition, at best limited attempts have been made to interpret the results of the new network metrics using well-known, established facts in seismology. Many of the claimed novel features are probably very well understood -- they are mostly related to scale-invariance and clustering of seismicity, facts documented for decades. The authors should always strive to show that the new metrics that they propose give results that cannot be explained by the standard laws in statistical seismology. In this goal, there are well-defined benchmark models that incorporate these laws, that generate synthetic catalogs on which the new metrics can be tested and compared.

A few exceptions are worth mentioning. Motivated by the long-standing and unresolved debate over ``aftershock" identifcation, \citet{BaiesiPaczuski2004, BaiesiPaczuski2005} and \citet{Baiesi2006} provided a new metric for the correlations between earthquakes based on the space-time-magnitude nearest-neighbor distance between earthquakes. The authors compared their results with known statistical laws in seismology and with the predictions of the ETAS model, actually confirming both. While no new law has been unearthed here, such 
efforts are valuable to validate known laws and continue to test the possible limits. A recent preprint \citep{Zaliapin-et-al2007} extended their study and investigated the theoretical properties of the metric and its ability to decluster catalogs (i.e., separate mainshocks from aftershocks). They conclude that aftershocks defined from this metric seem to be different from the rest of earthquakes. It will be interesting to see head-to-head comparisons with current state-of-the-art probabilistic declustering techniques that are based on empirical statistical laws and likelihood estimation \citep{Kagan1991a, Zhuang-et-al2002, Zhuang-et-al2004}.

\section{~Future Directions}

The study of the statistical physics of earthquake remains wide-open with many significant discoveries to be made. The promise of a holistic approach -- one that emphasizes the interactions between earthquakes and faults -- is to be able to neglect some of the exceedingly complicated micro-physics when attempting to understand the large scale patterns of seismicity. The marriage between this conceptual approach, based on the successes of statistical physics, and seismology thus remains a highly important domain of research. In particular, statistical seismology needs to evolve into a genuine
physically-based statistical physics of earthquakes.

The question of renormalizability of models of earthquake occurrence and the role of small earthquakes in the organization of seismicity is likely to remain an important topic. It connects with the problem of foreshocks and the predictability of large events from small ones and therefore has real and immediate practical applications as well as physical implications. 

More detailed and rigorous empirical studies of the frequency-size statistics of earthquake seismic moments and how they relate to seismo-tectonic conditions are needed in order to help settle the controversy over the power-law versus the characteristic event regime, and the role of regime-switching and universality. 

Spatially extended, dynamically evolving fault networks and their role in the generation of earthquakes are mostly ignored in the statistical physics approach to seismicity. Akin to the filaments in turbulence, these may provide key insights into the spatio-temporal organization of earthquakes. Novel methods
combining information from seismology to faulting will be required (e.g., \citep{SornetteA1,SornetteA2,SornetteGeilo,Gorshkov-et-al2003,OuillonDucorSor08}) to build a real understanding of the self-organization of the chicken-and-egg structures that earthquakes-faults constitute. Furthermore, a true physical approach requires understanding the spatio-temporal evolution of stresses, their role in earthquake nucleation via thermally activated processes, in the rupture propagation and in the physics of arrest, both involved in the generation of complex stress fields.

The important debate regarding statistical physics approaches to seismicity would benefit significantly from two points. Firstly, earthquake catalogs contain data uncertainties, biases and subtle incompleteness issues. Investigating their influence on the results of data analyses inspired by statistical physics increases the relevance of the results. Secondly, the authors should make links with the literature on statistical seismology which deals with similar questions. It is their task to show that the new metrics that they propose give results that cannot be explained by the standard laws in statistical seismology. For this, there are well-defined benchmark models that incorporate these laws, that generate synthetic catalogs on which the new metrics can be tested.

\end{document}